# Induced nonlinear phase shift of forward volume spin waves in magnetic films and one-dimensional magnonic crystals

Alexey B. Ustinov[1], Roman V. Haponchyk[1], Anton P. Burovikhin[1], Mitsuteru Inoue[2], Taichi Goto[2]

*1. Department of Physical Electronics and Technology, St. Petersburg Electrotechnical University, St. Petersburg, 197376 Russia*

*2. Research Institute of Electrical Communication, Tohoku University, 2-1-1 Katahira, Aoba, Sendai, Miyagi 980-8577, Japan*

ABSTRACT

A differential phase shift of a low-power spin wave (SW) induced by a high-power pumping wave co-propagating at different frequencies in perpendicularly magnetized magnetic films has been studied. We find that this effect for forward volume SWs propagating in yttrium iron garnet (YIG) films is stronger than that for surface SWs propagating in tangentially magnetized films. The results show that the induced nonlinear phase shift up to 180° takes place for pumping wave power of a few milliwatts. The phenomenon paves the way for fast and energy-efficient control of one-dimensional magnon transport.

Corresponding author: Alexey B. Ustinov

e-mail: ustinov_rus@yahoo.com

## I. Introduction

Recent advances in spin wave computing promise the future development of different magnonic devices [1]. The elaboration of linear and nonlinear spin wave (SW) interferometers of Mach-Zehnder type [2-4] resulted in designing magnonic logic gates based on linear SWs [5-7] and nonlinear SWs [8,9]. Fast and energy efficient control of magnon transport is highly demanded for further progress in this field.

Various methods can be used for the electronic tuning of the characteristics of magnonic devices. A variation of bias magnetic field is the widely used way [10,11]. One more efficient way is a local control of the bias field by electric current in the conducting wire which was used in the pioneering work on SW logic gate [5]. It is also possible to implement the tuning with the utilization of the ferrite-ferroelectric [12-14], ferrite-piezoelectric [15,16], and ferrite-semiconductor [17] structures.

We show recently that the effect of induced nonlinear phase shift can be used for a phase control of the SWs propagating in magnetic films [18] and one-dimensional (1D) magnonic crystals (MCs) [19] which is required for developments of SW logic gates. Experiments were carried out with surface SWs propagating in tangentially magnetized films. Out-of-plane magnetization provides a forward volume spin wave (FVSW) and offers a host of promising features such as a voltage-controlled perpendicular magnetic anisotropy [20-22], an increased nonlinearity [2] and suitability for device integration due to in-plane isotropy [23,24].

The aim of this work is investigation of an induced nonlinear phase shift of FVSWs propagating in perpendicularly magnetized magnetic films and 1D magnonic crystals. The paper is organized as follows. Section II describes experimental setup and measurement procedure. Section III presents experimental and theoretical data for regular and periodic magnetic film waveguides. Section IV provides summary and conclusions.

## II. Experimental setup and measurement procedure

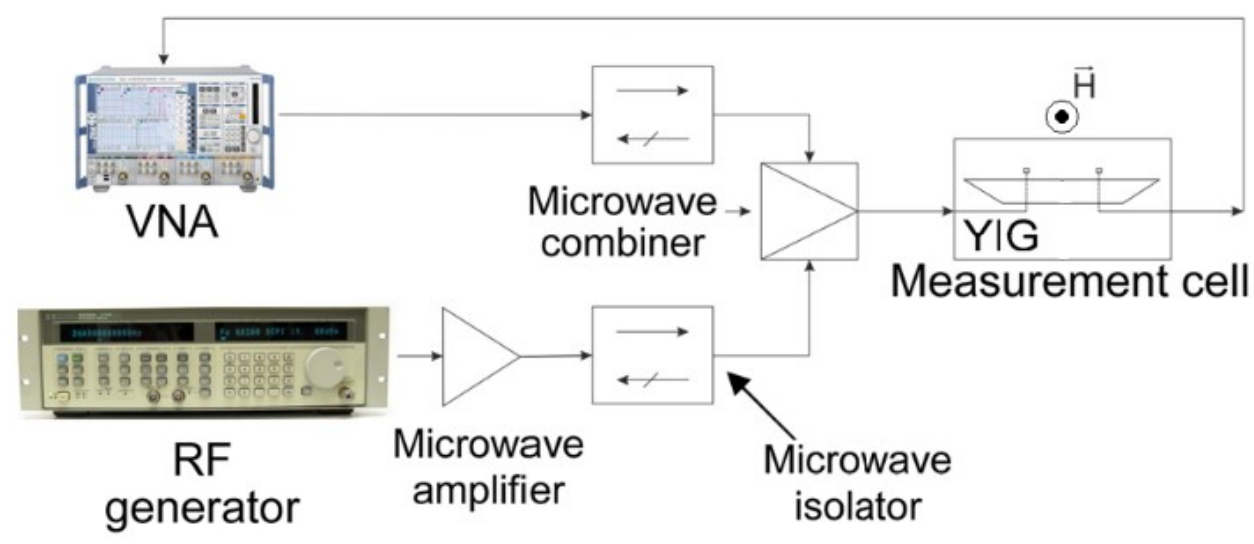

**Fig. 1.** Schematic diagram of the experimental setup.

The experimental setup is shown in Fig. 1. It consists of a vector network analyzer (VNA) Rohde & Schwarz ZVA 40, a radio frequency (RF) signal generator Agilent 83752A, a microwave amplifier Mini-Circuits ZVE-8G+, isolators, a combiner, a measurement cell, and a permanent magnet (not shown) providing perpendicular magnetization of the yttrium iron garnet (YIG) film. The setup is working as follows. A low-power continuous microwave signal called as "probe signal" was taken from the first port of the VNA. A high-power pulsed microwave signal called as "pumping signal" was generated by the RF signal generator. These signals were supplied simultaneously to the measurement cell. The cell has a form of the SW nonlinear phase shifter [3,18,19]. The device output signal was supplied to the second port of the VNA. A bias magnetic field was produced with the permanent magnet with adjustable field magnitude.

The characterization of magnetic films and magnonic crystals is carried out by measuring their amplitude-frequency characteristic (AFC). For this purpose we connected the VNA directly to the input and output ports of the measurement cell. The probe microwave signal power was 50 μW.

The experimental investigation of the induced nonlinear phase shift is carried out using the phase-vs-time measurement technique [18]. A low-power probe signal with an operating frequency $f_1$ and a power $P_1$ as well as high-power pump pulses with a carrier frequency $f_2$ and a power $P_2$ were supplied simultaneously to the measurement cell. As a result, we obtained phase versus time profiles with the VNA working in *the time domain measurement regime*. Typical traces measured for different values of pump pulse power are shown in Fig. 2. As is seen from the profiles, when the pumping signal is switched on, the phase of the low-power probe signal is changed. The induced nonlinear phase shift $\Delta\varphi_{1INL}$ was measured as the difference in the levels corresponding to 'switched on' and 'switched off' pumping signal regimes.

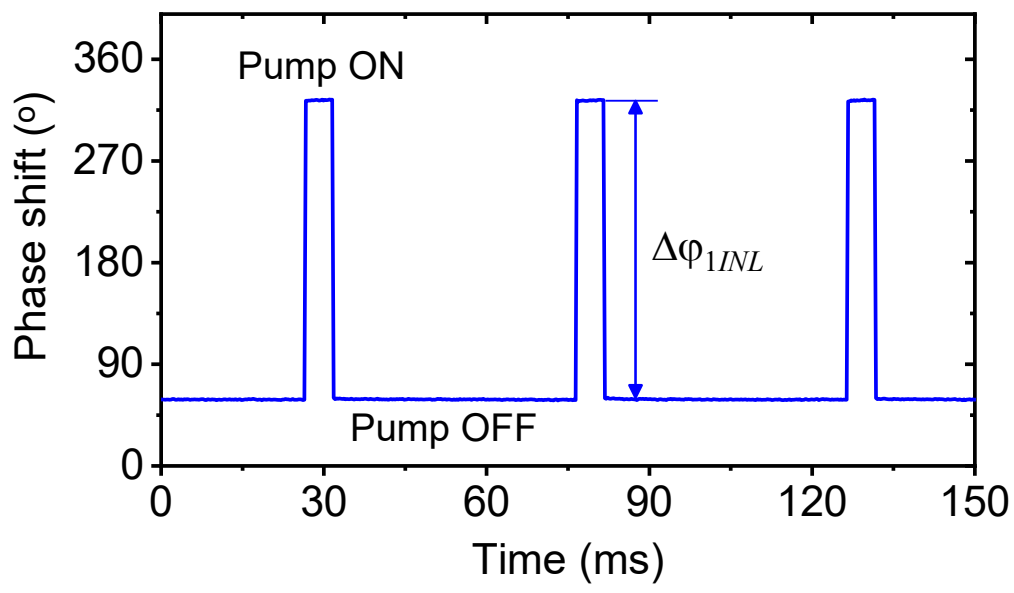

**Fig. 2.** Typical phase versus time characteristic measured with the VNA for pump power of 4.17 mW.

## III. Results and discussion

### A. Regular magnetic film waveguide

Fig. 3 shows a schematic of the measurement cell, which comprises a magnetic film waveguide integrated with spin-wave (SW) excitation and reception structures. Two 35-mm-long, 2-mm-wide single-crystal yttrium iron garnet (YIG) film waveguides, with thicknesses $L$ of 5.7 μm and 13.6 μm, were used in the experiments. The films were grown on 0.5-mm-thick gadolinium gallium garnet (GGG) substrates with liquid-phase epitaxy. The saturation magnetizations $4\pi M_S$ were 1825 G for the 5.7-μm film and 1905 G for the 13.6-μm film. Both films exhibited a narrow ferromagnetic resonance (FMR) linewidth $\Delta H$ of approximately 0.5 Oe at 5 GHz. A bias magnetic field of 2960 Oe was applied along the x-axis, as shown in Fig. 3. Spin waves are excited and detected using short-circuited microstrip antennas separated by 5 mm. The antennas have a width of 50 μm and a length of 2 mm. They were placed directly on the surface of the YIG film. Thus, from a technical point of view, the measurement cell represents a spin-wave nonlinear phase shifter [3].

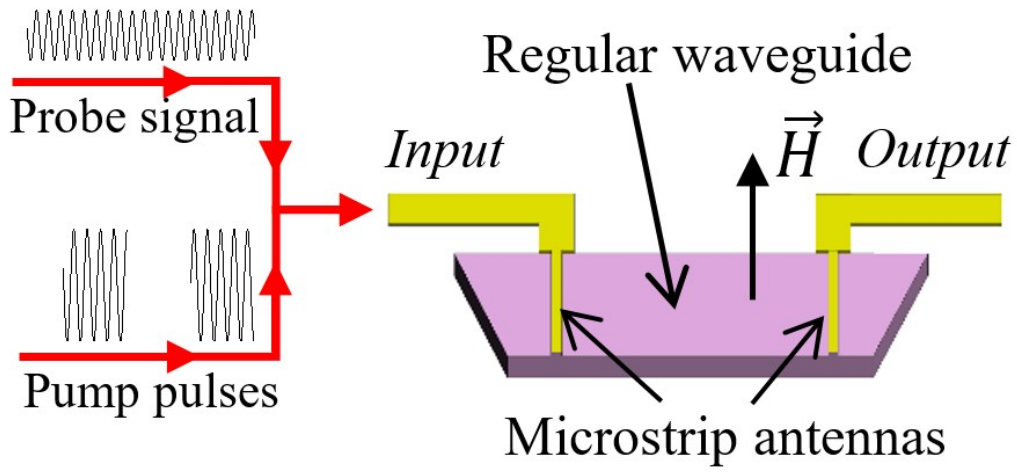

**Fig. 3.** Schematic diagram of the magnetic film experimental structure.

The AFCs of the measurement cells are presented in Fig. 4. As is seen, the characteristics have usual shape corresponding to spin-wave phase shifter based on FVSWs (see e.g. [25,26]).

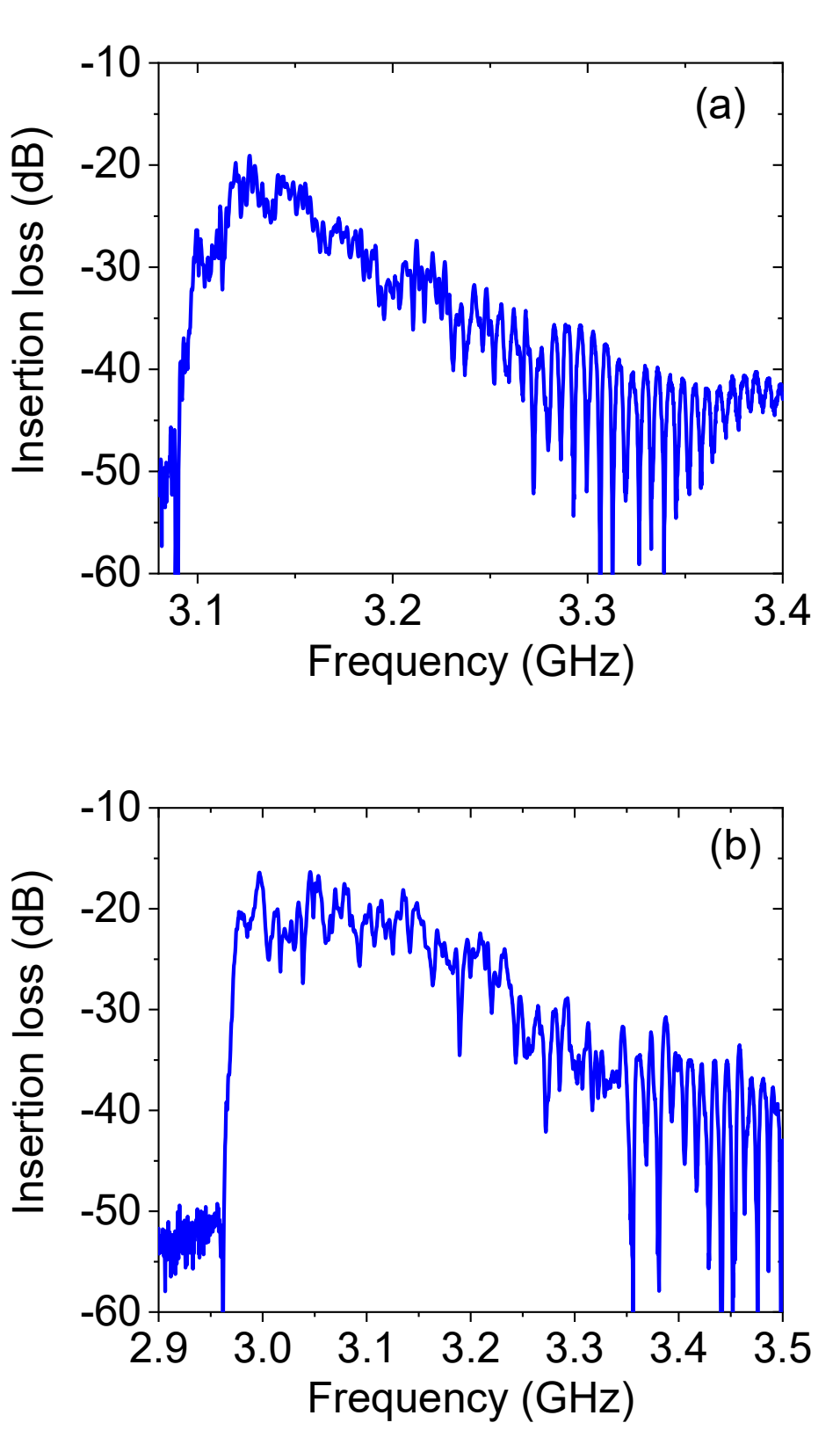

**Fig. 4.** AFCs of the measurement cells based on YIG films with thicknesses of 5.7 μm (a) and 13.6 μm (b).

The induced nonlinear phase shift characteristics are presented in Figures 5 and 6. Experimental data are shown by symbols, curves depict theoretical characteristics obtained as a result of numerical modelling. The induced nonlinear phase shift was calculated using the approach described in Ref. [18]. In consistent with this approach the nonlinear phase shift of the probe SW depends on its amplitude as well as on the amplitude of the pumping SW

$$\varphi_{1NL}(u_1, u_2) = \varphi_{1SNL}(u_1) + \varphi_{1INL}(u_2), \tag{1}$$

where $\varphi_{1SNL}$ and $\varphi_{1INL}$ are the self and the induced nonlinear phase shifts, respectively, $u_1$ and $u_2$ are the dimensionless amplitudes of the probe and pump SWs, respectively. The values of $\varphi_{1SNL}$ one can calculate with a theory presented in Ref. 27. For a small value of the operating SW amplitude the $\varphi_{1SNL} \approx 0$ whereas the induced nonlinear phase shift can be calculated numerically with the following expression

$$\varphi_{1INL}(y) = -\frac{N_{12}}{V_{g1}} \int_0^y u_{\omega 2}^2(y) dy, \tag{2}$$

where $V_{g1} = \partial\omega_1 / \partial k_1$ is the group velocity, $N_{12} = \partial\omega_1 / \partial |u_2|^2$ is the cubic nonlinear induced interaction coefficient and $u_{\omega 2}(y)$ is the amplitude profile of the pumping wave. This profile was calculated using nonlinear damping theory developed in Ref. 28. According to this theory, we expressed the total damping decrement of the pumping spin wave as $\omega_{r2} = \eta_2 + \nu_2 |u_2|^2 + \varsigma_2 |u_2|^4$, where $\eta_2$ is the linear damping parameter, $\nu_2$ and $\varsigma_2$ are the cubic and quintic nonlinear damping parameters,

respectively. A value of $\eta_2$ was determined from direct measurement of $\Delta H$ whereas the values of the nonlinear damping parameters $\nu_2$ and $\varsigma_2$ were determined from theoretical analysis of the experimental results.

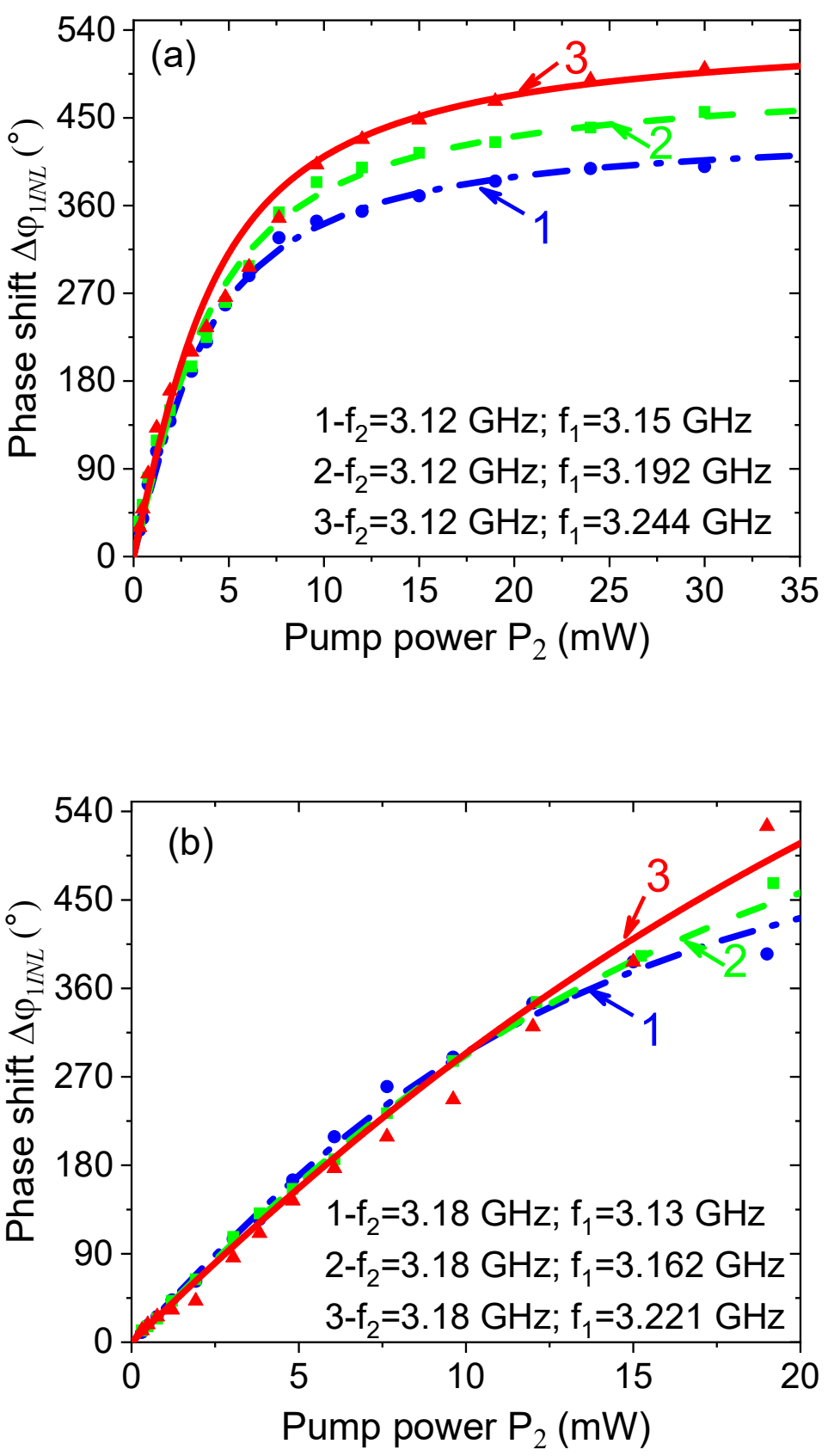

**Fig. 5.** The induced nonlinear phase shift of the probe FVSWs measured for the 5.7-µm-thick YIG film. Panels (a) and (b) corresponds to $f_2$ = 3.12 GHz and $f_2$ = 3.18 GHz, respectively. Experimental data are shown with symbols, while calculation results are shown with lines.

Figure 5(a) presents the measured nonlinear phase shift $\Delta\varphi_{1INL}$ as a function of the pumping wave power $P_2$ at a fixed pumping frequency $f_2$ = 3.12 GHz for several probe signal frequencies $f_1$. The increase in $\Delta\varphi_{1INL}$ is nearly linear with respect to $P_2$ up to approximately 5 mW. Above this power, $\Delta\varphi_{1INL}$ saturates, which is attributed to the nonlinear damping of the high-power pumping SW. Furthermore, $\Delta\varphi_{1INL}$ increases with the operating wave frequency $f_1$. This trend arises from the reduction in the operating SW's group velocity at higher frequencies (see Eq. (2)). One can also see from the presented results that a differential phase shift $\Delta\varphi_{1INL}$ of 180° accumulates in the probe signal at a pumping power $P_2$ as low as 2.1-2.8 mW, depending on the probe signal frequency.

Figure 5(b) shows data for a pumping frequency of $f_2$ = 3.18 GHz under otherwise identical conditions. In this case, $\Delta\varphi_{1INL}$ tends to saturation from the pump power of about 10 mW. At the same time, evident saturation in the characteristic behavior was not observed. This indicates about weak fifth-order nonlinearity.

Similar behavior was observed for the 13.6-μm-thick YIG film (see Fig. 6). The induced nonlinear phase shift reaches more than 180° at several tens of milliwatts of pump power. This thicker film demonstrate weaker nonlinearity due to lower intensity per unit volume and higher group velocity of the pumping SW. Table 1 provides the parameters $v_2$ and $\varsigma_2$ for FWSWs.

**Table 1.** The list of parameters $v_2$ and $\varsigma_2$

| $L$ (μm) | $f_1$ (GHz) | $f_2$ (GHz) | $v_2$ (1/ns) | $\varsigma_2$ (1/ps) |
|---|---|---|---|---|
| 5.7 | 3.150 | 3.120 | 0.2 | 1.05 |
| 5.7 | 3.192 | 3.120 | 0.4 | 0.8 |
| 5.7 | 3.244 | 3.120 | 0.4 | 0.75 |
| 5.7 | 3.130 | 3.180 | 0.01 | 0.45 |
| 5.7 | 3.162 | 3.180 | 0.2 | 0.3 |
| 5.7 | 3.221 | 3.180 | 0.5 | 0.1 |
| 13.6 | 3.105 | 3.050 | 0.1 | 1.9 |
| 13.6 | 3.290 | 3.050 | 0.4 | 3 |
| 13.6 | 3.408 | 3.050 | 0.1 | 2.7 |
| 13.6 | 2.996 | 3.150 | 0.4 | 0 |
| 13.6 | 3.175 | 3.150 | 1.5 | 0 |

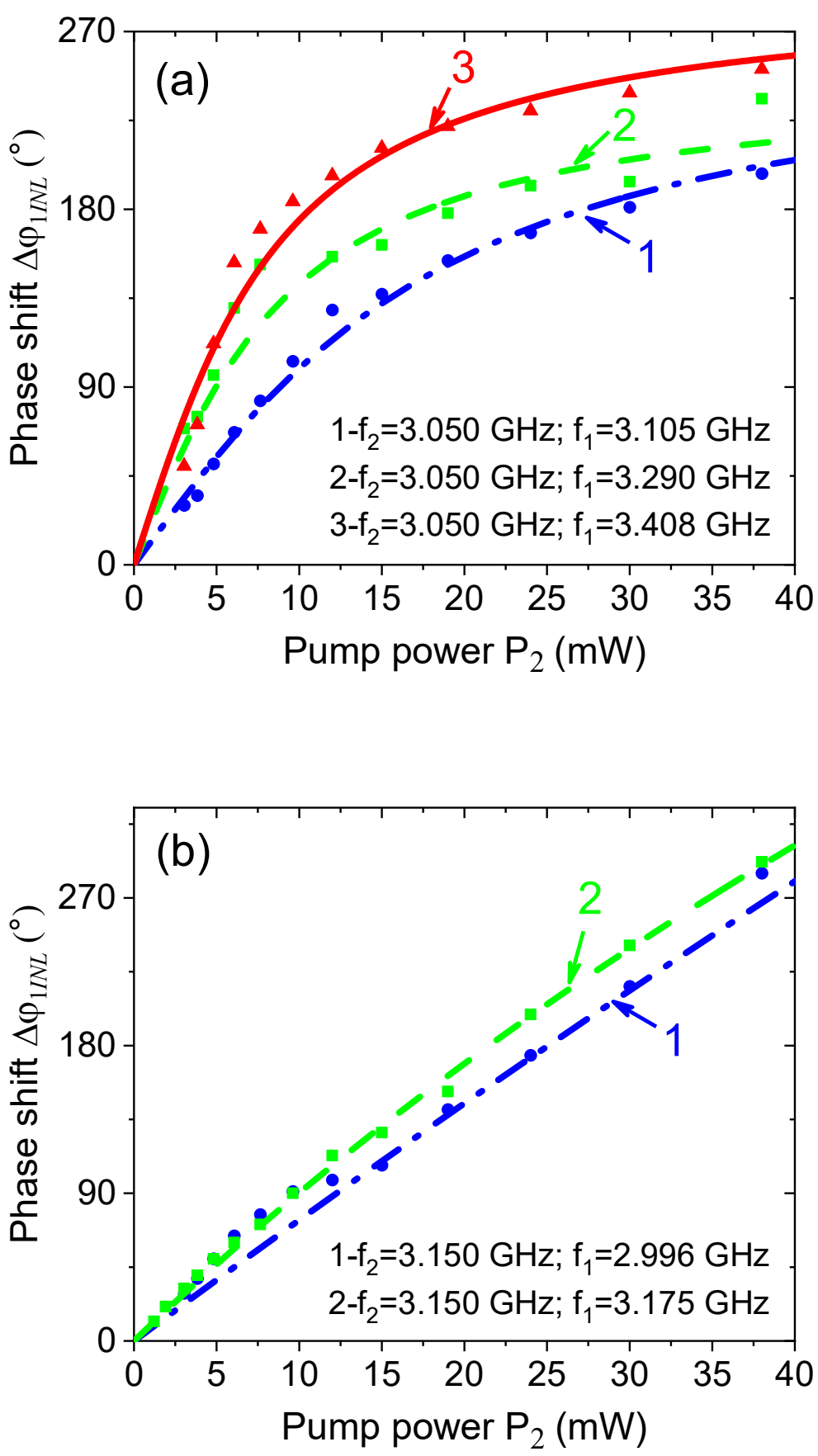

**Fig. 6.** The induced nonlinear phase shift of the probe FVSWs measured for the 13.6-μm-thick YIG film. Panels (a) and (b) corresponds to $f_2$ = 3.05 GHz and $f_2$ = 3.15 GHz, respectively. Experimental data are shown with symbols, while calculation results are shown with lines.

### B. Periodic magnetic film waveguides – 1D MCs

Figure 7(a) shows schematically the experimental 1D MC as well as the SW excitation and reception structure. The periodic structure is fabricated from a 5.7-μm-thick YIG film, which is different from that described in the previous subsection. The film has a saturation magnetization of 1400 G and $\Delta H \approx 0.6$ Oe. Ten grooves with a depth of 0.2 μm and a period of 300 μm are etched on the film surface, with each groove having a width of 100 μm. Microstrip antennas with a length of 2 mm and a width of 50 μm are used to excite and receive the FVSWs. The antennas are positioned directly on the YIG film near the edges of the MC with a distance of 3 mm. A bias magnetic field of 2495 Oe is applied perpendicular to the surface of the MC, as shown in Fig. 7(a).

Figure 7(b) shows experimental and theoretical results of the frequency response of the magnonic nonlinear phase shifter. Numerical modeling was carried out using T-matrix's method [29] and SW excitation theory [30]. It is seen, that this theoretical approach is suitable for describing the AFC of 1D MC and makes it possible to determine theoretically the number, depth, and frequency of the magnonic band gaps.

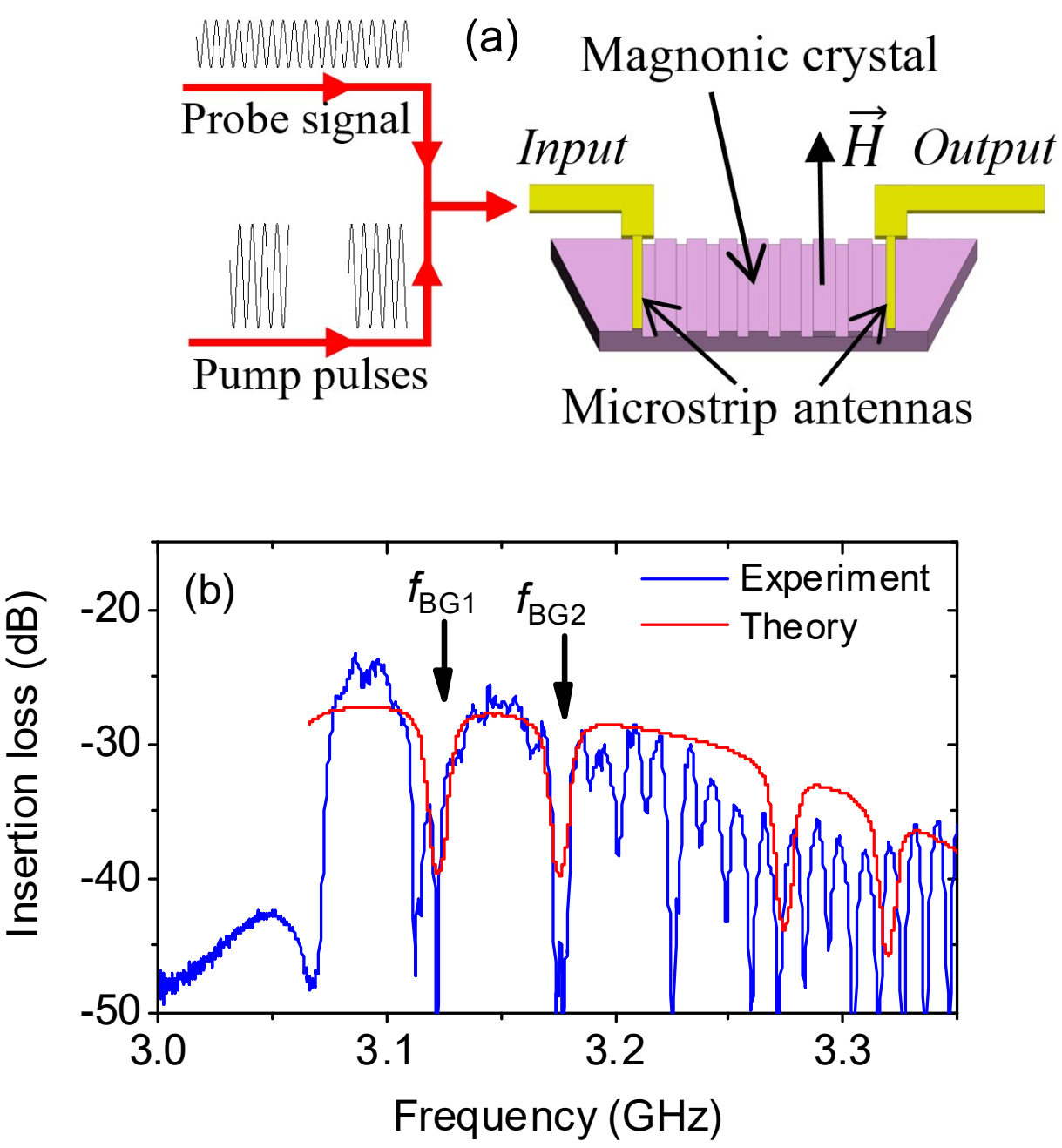

**Fig. 7.** Schematic diagram of the experimental setup using a 1D MC(a), and its amplitude-frequency characteristic (b).

The experimental dependences of the induced nonlinear phase shift $\Delta\varphi_{1INL}$ of the probe SW as a function of the pump signal power $P_2$ are shown by symbols in Fig. 8. The probe SW excitation power is 10 μW. Frequencies of the pump signal are chosen both outside and inside the magnonic band gaps. The results show that when the pumping frequency $f_2$ is set to the middle of a band gap, the induced nonlinear phase shift $\Delta\varphi_{1INL}$ is decreased. This is explained by enhanced attenuation of the forward pumping SW caused by power transfer to the reflected wave. A theoretical approach considering this effect can be found in our previous work [19]. The numerically simulated characteristics of the induced nonlinear phase shift are shown in Fig. 8 by solid lines. By comparing the data, the values of the nonlinear damping parameters are determined and are listed in Table 2. Good agreement between experimental and theoretical data can be seen.

Figure 9 shows the induced nonlinear phase shift of the probe signal measured for various pump microwave signal frequencies at a constant probe signal frequency $f_1$ of 3.208 GHz. The experimental data show that the induced nonlinear phase shift decreases when the pump frequencies are at the middle of the band gaps (shown by arrows in Fig. 9). This provides further confirmation that additional attenuation of the pump SW occurs due to its coupling with the reflected wave.

**Table 2.** The list of parameters $v_2$ and $\varsigma_2$ for a 1D MC.

| $f_1$ (GHz) | $f_2$ (GHz) | $v_2$ (1/ns) | $\varsigma_2$ (1/ps) |
|---|---|---|---|
| 3.153 | 3.097 | 0.1 | 0 |
| 3.153 | 3.122 | 0.01 | 0.49 |
| 3.153 | 3.167 | 1 | 0.3 |
| 3.153 | 3.177 | 0.85 | 0 |
| 3.153 | 3.192 | 0.7 | 0 |
| 3.208 | 3.097 | 1.2 | 0 |
| 3.208 | 3.122 | 0.1 | 0.35 |
| 3.208 | 3.167 | 0.1 | 0 |
| 3.208 | 3.177 | 0.1 | 0.02 |
| 3.208 | 3.192 | 0.01 | 0.14 |

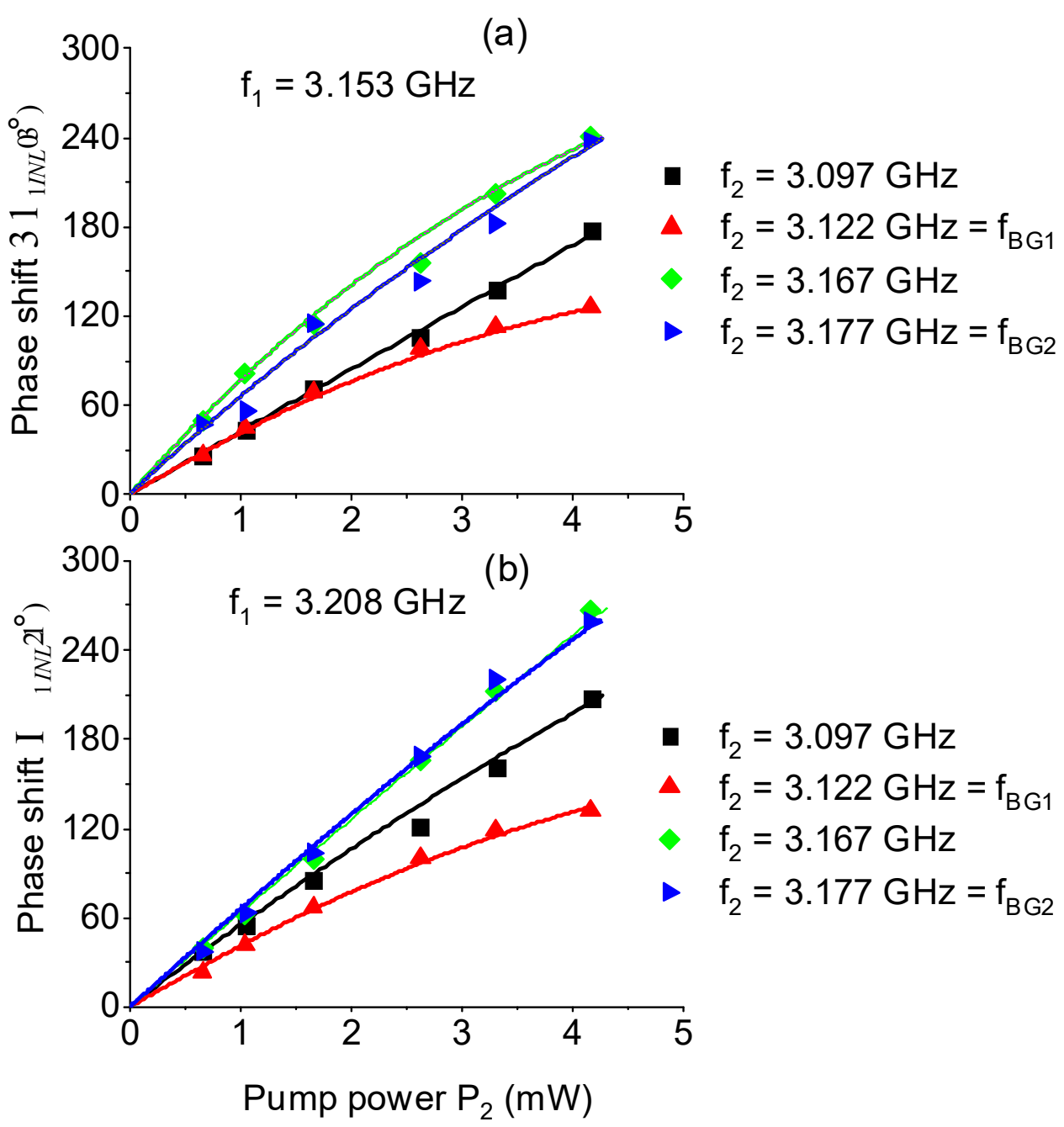

**Fig. 8.** The induced nonlinear phase shift of the probe FVSWs measured for MC. Panels (a) and (b) shows data for different values of $f_1$ as indicated. Experimental data are shown with symbols, while calculation results are shown with lines.

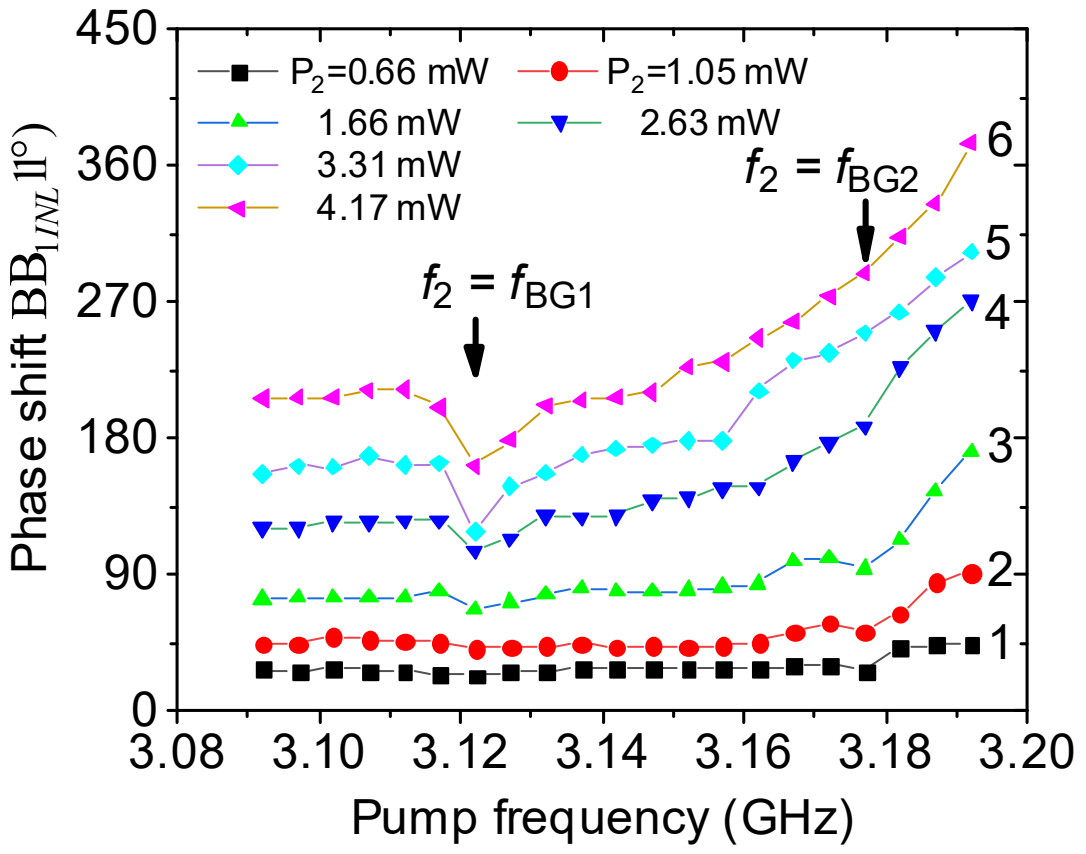

**Fig. 9.** The induced nonlinear phase shift of the probe FVSWs in MC at the frequency $f_1$ = 3.208 GHz as a function of pump frequency $f_2$ measured for different pump powers $P_2$.

## IV. Conclusion

Our research show that the induced nonlinear phase shift of FVSWs exceeds 180° for pump powers of a few milliwatts. This is drastically smaller than that required for surface SWs [18,19]. Thus, FVSW may have advantages for a variety of practical applications requiring low power consumption. For example, the nonlinear phase shift may be used for development of magnonic logic circuits [1] and reservoir computing devices [31]. The experiments were carried out for YIG films having microscale thickness. Further reduction in pump power could be achieved by using YIG films with thicknesses of tens or hundreds of nanometers [32]. Therefore, the investigated effect provides a pathway to fast and energy-efficient control of one-dimensional magnon transport.

## Acknowledgement

Authors of this manuscript acknowledge Dr. Nikolai Kuznetsov for the data presented in Figures 5 and 6. The raw data were obtained by him during the 2019–2020 academic year under the supervision of Prof. Alexey Ustinov and Prof. Erkki Lähderanta. Some of the data were also used in Dr. Kuznetsov's Master's thesis [33] and have not been published elsewhere.

This work was partly supported by the Ministry of Science and Higher Education of the Russian Federation (Grant No. FSEE-2025-0008).

## Conflict of interest

The authors have no conflicts to disclose.

## Data availability

The data that support the findings of this study are available from the corresponding author upon reasonable request.